\documentclass[usenatbib]{article}  
\usepackage{amsmath,amsfonts,amssymb}
\usepackage{graphicx}
\usepackage{fancyhdr}
\newcommand{\be}{\begin{equation}}
\newcommand{\ee}{\end{equation}}
\newcommand{\lo}{{\cal L}_0}

\title{Monitoring daytime and nighttime optical turbulence profiles with the PML instrument} 

\author{Eric Aristidi$^1$, Aziz Ziad$^1$, Yan Fant\'e\"{\i}-Caujolle$^1$, Julien Chab\'e$^2$, Christophe Giordano$^1$,\\ Catherine Renaud$^1$, Henri Lant\'eri$^1$}
\date{\small $^1$: UMR 7293, Lagrange, Universit\'e de Nice-Sophia Antipolis, CNRS, OCA, Parc Valrose F-06108 Nice Cedex 2, France\\ $^2$: Universit\'e C\^ote d'Azur, Observatoire de la C\^ote d'Azur, CNRS, IRD, G\'eoazur, 2130 route de l'Observatoire, 06460 Caussols, France\\ \ \\
\it AO4ELT6, Qu\'ebec, June 6-14, 2019}

\begin{document} 
  
  \maketitle 

\begin{abstract}
The Profiler of Moon Limb is a recent instrument dedicated to the monitoring of optical turbulence profile of the atmosphere. Fluctuations of the Moon or the Sun limb allow to evaluate the index refraction structure constant $C_n^2(h)$ and the wavefront coherence outer scale $\lo(h)$ as a function of the altitude $h$. The atmosphere is split into 33 layers with an altitude resolution varying from 100m (at the ground) to 2km (in the upper atmosphere). Profiles are obtained every 3mn during daytime and nighttime. We report last advances on the instrument and present some results obtained at the Plateau de Calern (France).
\end{abstract}



\section{INTRODUCTION}
\label{par:intro}  
The Profiler of Moon Limb (PML) is an instrument dedicated to the monitoring of optical turbulence profile of the atmosphere. It allows measurement of the vertical distribution of the refractive index structure constant $C_n^2(h)$ as a function of the altitude $h$. It also provides values of the integrated seeing and isoplanatic angle. The principle of PML is to observe the Moon or the Sun limb: these extended objects act as a continuum of angular separations (2 points on the limb are considered as a double star) allowing the scan of the atmosphere with fine resolution (100m near the ground). One asset of PML is the access to daytime profiles which is difficult with other profiler instruments using star based optical techniques such as the Generalized Scidar\cite{Fuchs98} or the Multi-Aperture Scintillation Sensor (MASS)\cite{Kornilov07}.

PML was developed in the early 2010s. It is based on a small commercial telescope (diameter 40cm) equiped with a 2 sub-apertures mask. Two copies of the instrument were built: one ``winterized'' version was sent to Dome C in Antarctica and gave the first $C_n^2(h)$ profiles in 2011\cite{Ziad13}. The other one, a lighter version designed for mid latitudes sites, gave also its first results in 2011 at the South African Large Telescope (SALT)\cite{Catala16}. 

At the end of 2015, the ``antarctic'' PML was installed at the  Plateau de Calern observatory (France, UAI code: 010, Latitude=43$^\circ$45$'$13$''$N, Longitude=06$^\circ$55$'$22$''$E). The telescope mount was placed on the top of a 1.5m~high concrete pillar, and protected by a 12~feet allsky dome. PML is now part of the Calern Atmospheric Turbulence Station (CATS)\cite{Chabe16, Ziad18}, who was developped as a site monitoring facility at Calern, in connexion with the laser telemetry MeO station\cite{Samain08}.  The other instruments composing CATS are the Generalized Differential Motion Monitor\cite{Aristidi19}, an All-Sky camera providing the cloud coverage during the night and a meteo station. To perform daytime observations, a solar pyranometer (Davis Instruments 6450) giving the solar irradiance was added to the meteo station and calibrated to provide an estimate of the cloud cover during the day.
More details on CATS are given in [\cite{Ziad18, Ziad19b}]. 

\begin{figure}
\includegraphics[width=15cm]{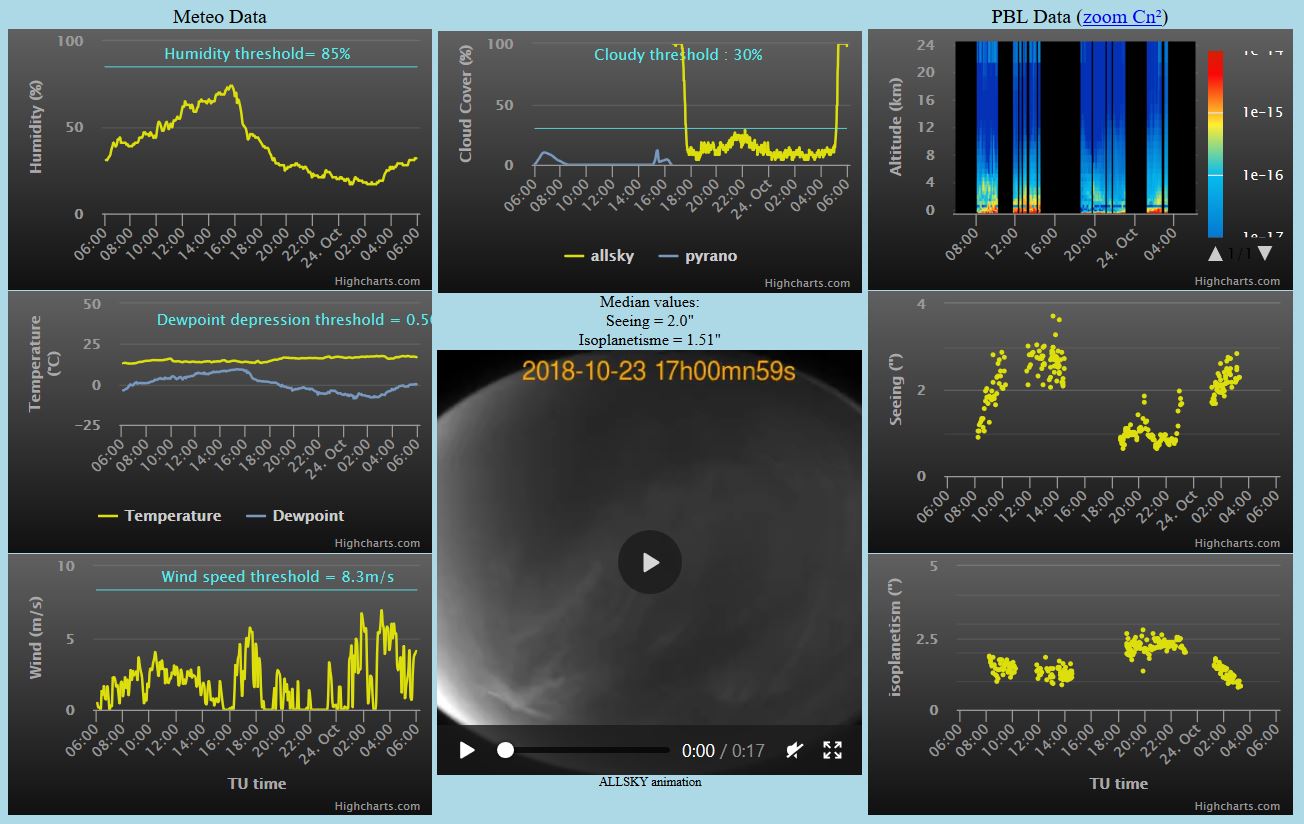}
\caption{PML website displaying real time measurements (the URL is {\tt cats.oca.eu}).}
\label{fig:srccopy}
\end{figure}

From 2016 to now on, a lot was done to make PML fully automatic, using informations from the meteo station, the All-Sky camera, and the pyranometer to decide whether or not it is possible to observe. Solar filters are placed automatically during daytime to switch from lunar to solar observations. As the automatisation progressed, the number of daily collected data became larger. The instrument is now running at its full capacity, providing a $C_n^2(h)$ profile as well as a seeing and isoplanatic angle values every 3mn, daytime and nighttime (providing that the Moon is visible). Data are accessible worldwide via the website {\tt cats.oca.eu}. A screen copy of the PML website is shown in Fig.~\ref{fig:srccopy}.

\section{PRINCIPLE OF THE INSTRUMENT}
\begin{figure}
\includegraphics[width=7cm]{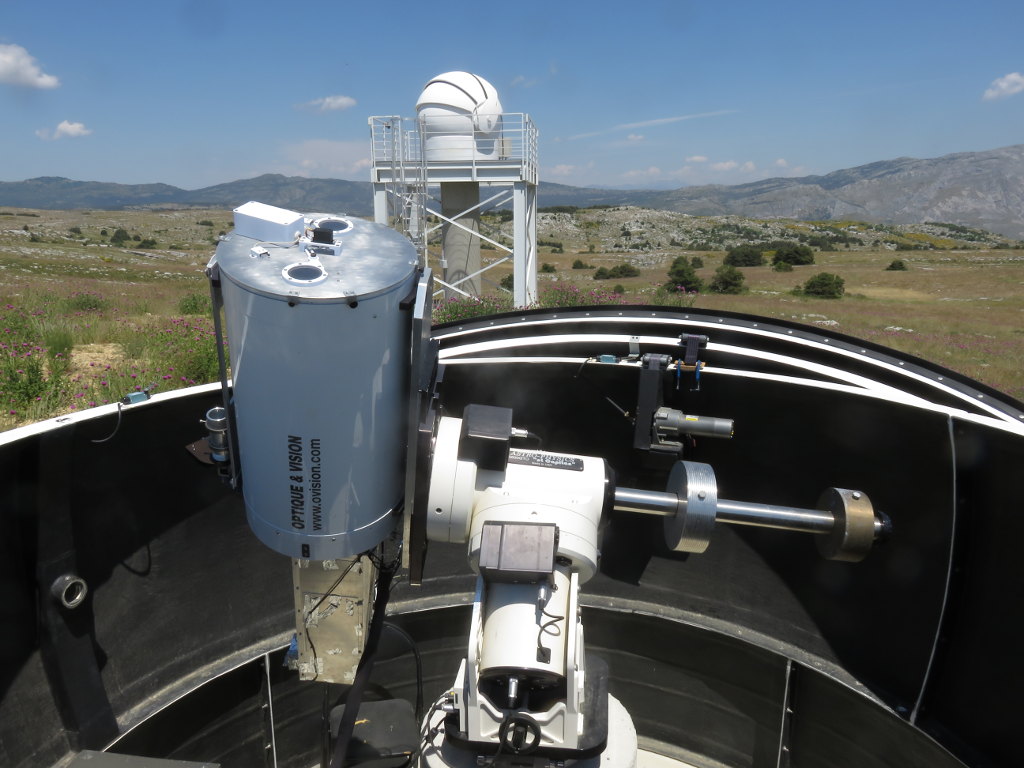}
\includegraphics[width=7cm]{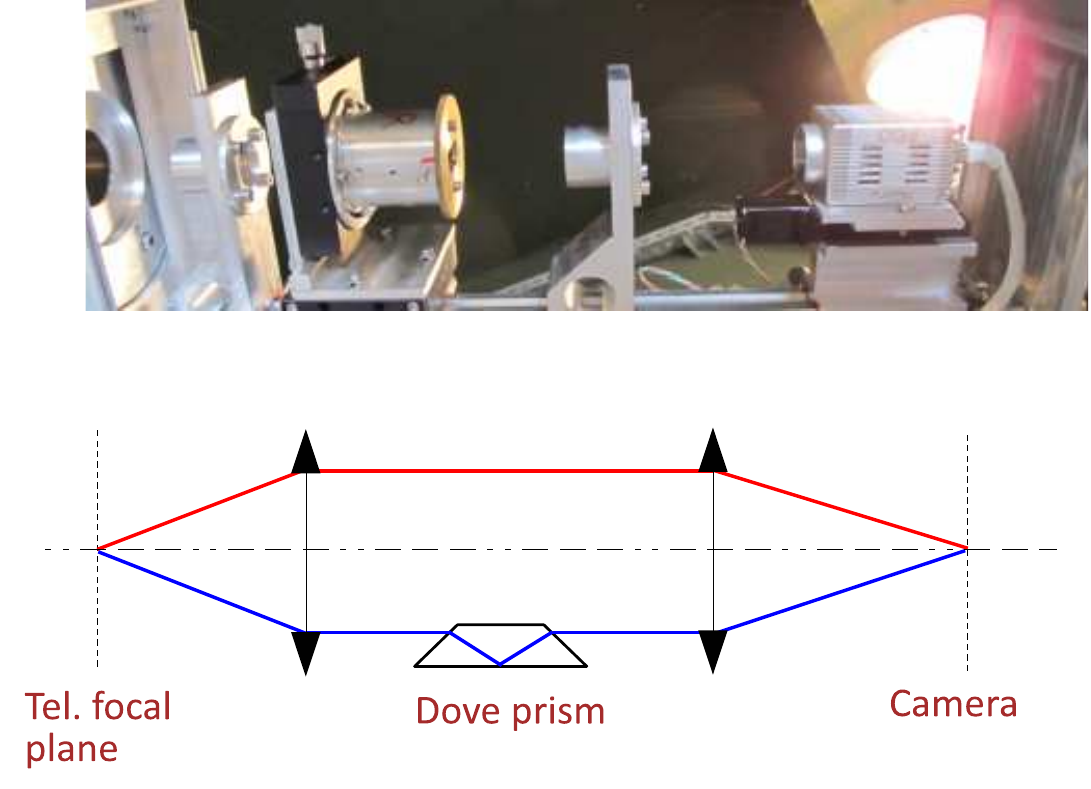}
\caption{Left: the PML instrument observing the Sun at the Calern observatory (in background one can see the tower of the GDIMM). Right: PML optical bench at the focus of the telescope. It is composed of two lenses creating an afocal beam: a Dove prism in introduced between the lenses to rotate one of the twin images of the Moon/Sun.}
\label{fig:photopml}
\end{figure}

The PML instrument is based on the measurement of the wavefront angle-of-arrival (AA) deduced from the motion of the Moon limb (or Sun edge) image. PML uses the differential method of the Differential Image Motion Monitor (DIMM)\cite{SarazinRoddier90} through two subapertures of diameter $D$ separated by a baseline $B$. Two images of the Moon or the Sun are produced at the focal plane of a small telescope, one being flipped by a Dove prism to produce mirror images of the Moon/Sun. Fig.~\ref{fig:photopml} shows the optical scheme of the PML bench, and a photo of the telescope in its dome at Calern.

PML is based on a small commercial Ritchey-Chr\'etien telescope of diameter 40cm mounted on an Astro-Physics AP3600 equatorial mount. The pupil mask has two sub-apertures of diameter $D=6$cm separated by a baseline $B=27$cm. The baseline is parallel to the declination axis. The camera is a PCO Pixelfly CCD with a matrix of 640$\times$480 pixels, working in the visible domain (the bandwidth FWHM is 320--630 nm). The sampling of 0.57~arcsec/pixel is a good compromise between the spatial resolution and the field of view (365~arcsec). Image cubes are composed of $N=1024$ frames with an exposure time of 5~milliseconds.

\begin{figure}
\includegraphics[width=13cm]{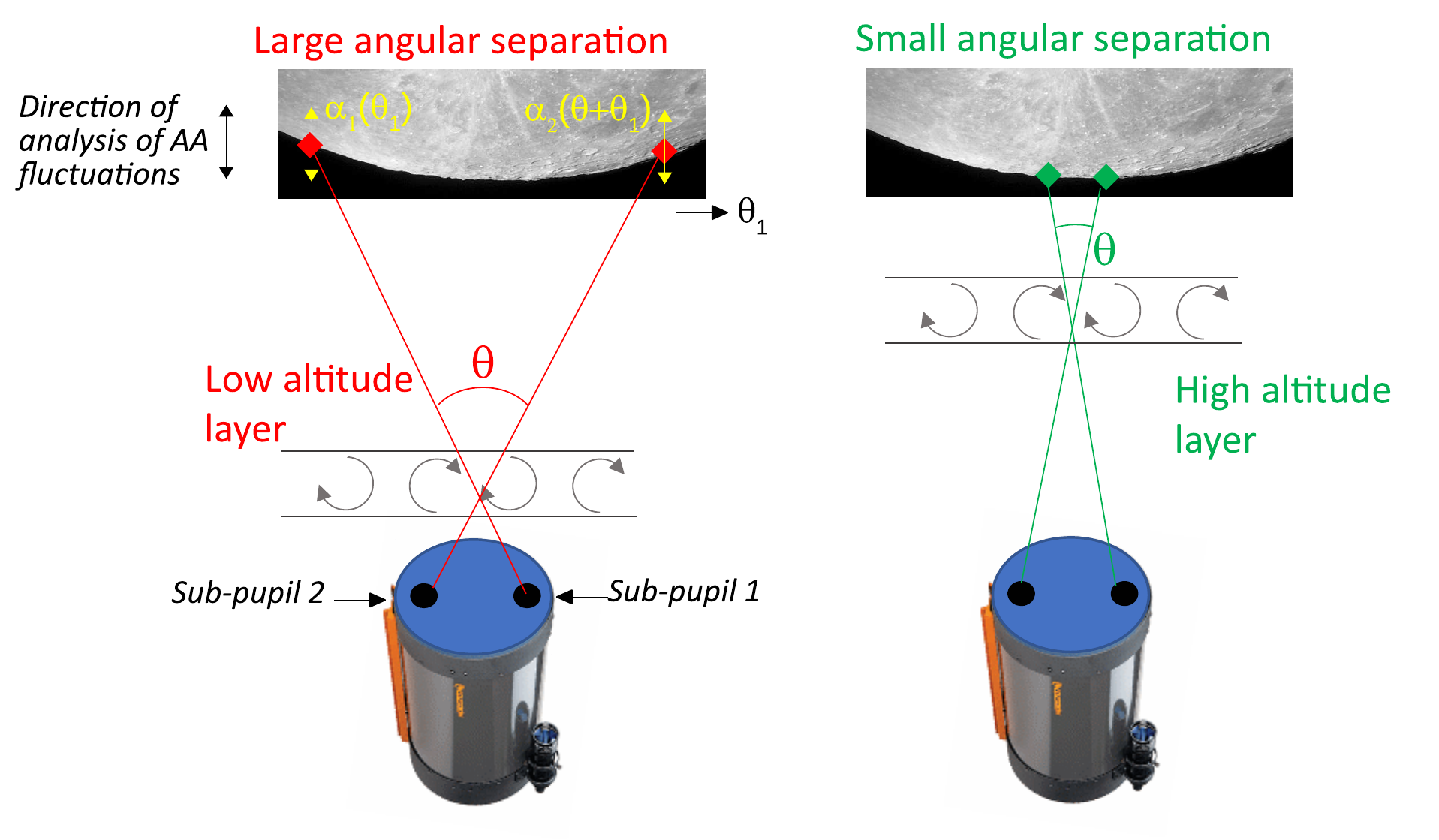}
\caption{Principle of PML measurement of the covariance of AA differential motion between two points on the Moon limb. Large (resp. small) angular separations give access to low (resp. high) altitude layers.}
\label{fig:pmlprinciple}
\end{figure}

The theoretical background of PML was presented in details elsewhere\cite{Ziad13, Catala16, Chabe19}. Here we will make a short summary of it. PML produces twin images of a Moon/Sun limb: we call $\alpha_{1|2}(\theta_1)$ the angular transverse position of a given point of the limb for the image produced by the sub-pupil 1 or 2 (see Fig.~\ref{fig:pmlprinciple}). $\theta_1$ is a coordinate measured along the limb (RA axis), while $\alpha$ is measured in the transverse direction (declination) to minimize the impact of vibrations due to the mount guiding.

The key quantity is the angular covariance of the differential motion of pair of points separated by an angle $\theta$ on the Moon or Sun limb:
\begin{equation}
C_{\Delta \alpha}(\theta)=\langle \langle \Delta\alpha(\theta_1) \; . \; \Delta\alpha(\theta_1+\theta)
\rangle_{\theta_1} \rangle_t
\end{equation}
where:
\begin{itemize}
\item $\Delta\alpha(\theta_1)=\alpha_2(\theta_1)-\alpha_1(\theta_1)$ is the angular difference of transverse positions of matching points between the two sub-images
\item $\theta$ is the angular separation between the two points of the couple. It varies from 0 to the total field of view of the instrument.
\item the brackets $\langle \rangle_{\theta_1}$ and $\langle \rangle_t$ stand respectively for spatial average over the position $\theta_1$ and for temporal average over the $N$ images of data cubes.
\end{itemize}
The theoretical expression of $C_{\Delta \alpha}(\theta)$ in the case of the Von-K\'arm\'an model, is given by\cite{Ziad13}
$$
C_{\Delta \alpha}(\theta)=\int_{h=0}^\infty C_n^2(h)\; \left[ 2 C_a (\theta h)-C_a(B-\theta h)-C_a(B+\theta h)\right]\; {\rm d}h
$$
with
$$
C_a(\rho)=1.2 \sec z\: \int_{f=0}^\infty f^3 \left(f^2+\lo^{-2}\right)^{-11/6}\; \left(J_0(2\pi\rho)+J_2(2\pi\rho)  \right)\left[2 \frac{J_1(\pi D f)}{\pi D f}\right]^2\; {\rm d}f
$$
where $z$ is the zenith distance of the Moon or the Sun, and $\lo$ the wavefront outer scale (taken equal to the standart value 20m in our data processing; the differential covariance is indeed not very sensitive to $\lo$).

The $C^2_n(h)$ profiles are retrieved by solving an inverse problem under non-negativity constraint, via the minimization of a least squares criterion between measured and modelled differential covariances. The algorithm is described in details in $[\cite{Chabe19}]$. Profiles are calculated on an altitude grid of 33 layers, from the ground up to $h=24$km. The resolution is $\Delta h=100$m in the first kilometer above the ground, then $\Delta h=500$m for $h\in[1250,4750]$m, then $\Delta h=1$km between 5 and 15km, and $\Delta h=2$km above. Uncertainties on $C_n^2(h)$ increase with the altitude, typical values are\cite{Chabe19}:

\parbox[t]{8cm}{
\begin{itemize}
\item near the ground : $\frac{\Delta C_n^2}{C_n^2} \lesssim 1\%$
\item 200m -- 1km : $\frac{\Delta C_n^2}{C_n^2} \simeq 3\%-5\%$
\end{itemize}}\ 
\parbox[t]{8cm}{
\begin{itemize}
\item 1km -- 10km : $\frac{\Delta C_n^2}{C_n^2} \simeq 15\%-25\%$
\item above 10km : $\frac{\Delta C_n^2}{C_n^2} \simeq 60\%-100\%$
\end{itemize}}

In addition to the $C^2_n(h)$ profile, PML has access to integrated parameters. The seeing $\epsilon$ can be obtained from the DIMM method\cite{SarazinRoddier90, Tokovinin02} through the differential variance $C_{\Delta \alpha}(0)$. Thanks to the large number of data points, this method gives a high accuracy on seeing values, of the order of $\frac{\Delta\epsilon}{\epsilon}= 0.3\%$\cite{Chabe19}. 

The isoplanatic angle $\theta_0$ is derived from the weighted integration of the $C^2_n(h)$ profile\cite{LoosHogge79}
\begin{equation}
\theta_0=0.528 \: \left[\left(\frac{2\pi}{\lambda}\right)^2 \int_0^\infty dh\, h^{5/3}\, C_n^2(h)
\right]^{-3/5}
\label{eq:isop}
\end{equation}
Currently we measure $\theta_0$ with a relative uncertainty of 15\% to 20\%\cite{Chabe19}. Another way to estimate $\theta_0$ from PML data without using the $C_n^2$ profile, was recently proposed\cite{Ziad19}. The technique is based on the correlation angle of the AA, calculated from AA structure functions of Moon/Sun limbs. It gives results similar to the present technique, and provides a way to cross check the consistency of the reconstructed $C^2_n(h)$ profiles.


\section{RESULTS}
\begin{figure}
\begin{center}
\includegraphics[width=8cm]{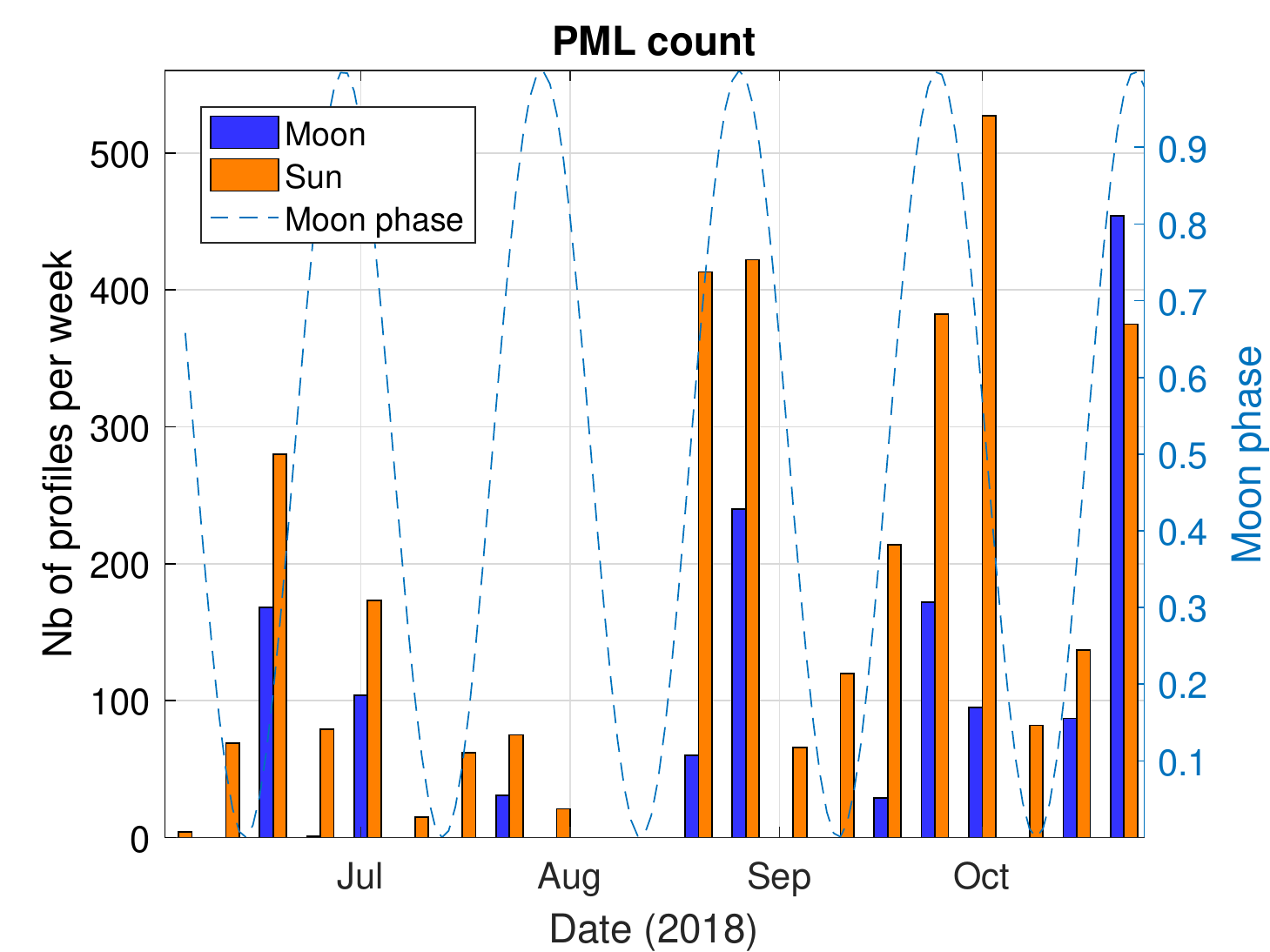}
\end{center}
\caption{Number of $C_n^2$ profiles obtained per week on the Sun and the Moon during the period June--October 2018. The dashed curve corresponds to the Moon phase (1=full Moon, 0=new Moon).}
\label{fig:nbdataweek}
\end{figure}

In this paper we will focus on data obtained between June and October 2018. A total of 4960 profiles were obtained during this period: 3520 on the Sun (during daytime) and 1440 on the Moon (nighttime). Figure~\ref{fig:nbdataweek} shows the number of profiles obtained for each week of the period. As one can see, we obtained a larger data set on the Sun, since the Moon is not always present in the sky (especially in Summer where the Moon has a low altitude above the horizon when it it close to the full Moon).

\subsection{Statistics of $C_n^2(h)$ profiles}
\begin{figure}
\begin{center}
\includegraphics[width=7cm]{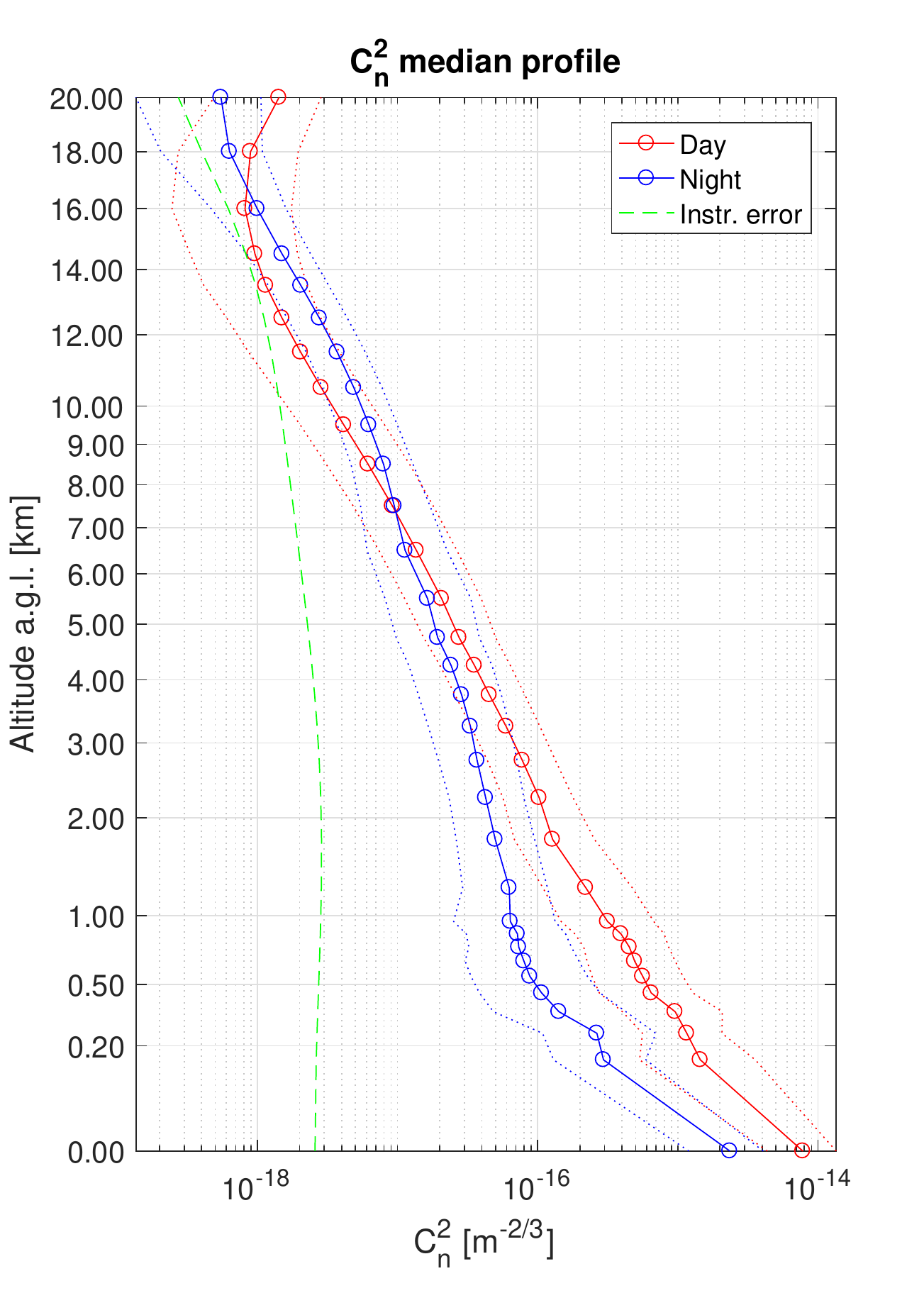}\ 
\includegraphics[width=7cm]{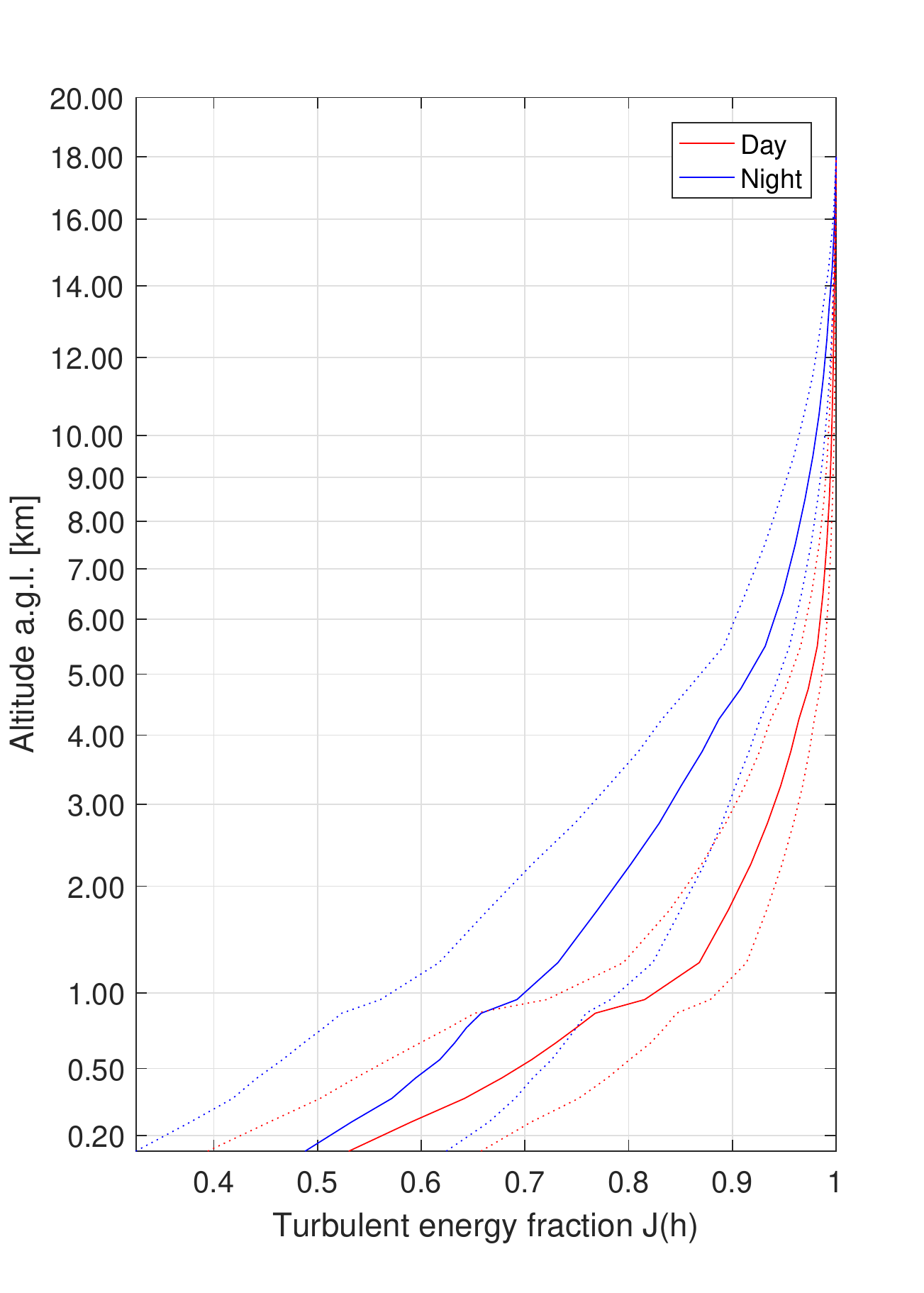}\\
\end{center}
\caption{Left: daytime and nighttime median $C_n^2$ profiles obtained from Sun and Moon limbs during the period June--October 2018. Dotted lines represent the first and third quartile. The green dashed line is the instrumental uncertainty (details on its calculation are given in [\cite{Chabe19}]). Right: turbulent energy fraction $J(h)$, defined in Eq.~\ref{eq:J(h)}, contained between the ground and the altitude $h$.}
\label{fig:cnmedian}
\end{figure}

Median profiles of $C_n^2(h)$, calculated on lunar and solar data, are displayed in Fig.~\ref{fig:cnmedian} (on the left). Unsurprisingly, they show that the turbulence is stronger during the day, especially in the first kilometer above the ground. At the altitude of $h=150$m, nighttime and daytime $C_n^2$ median values are respectively $3. 10^{-16} {\rm m}^{-2/3}$ and $1.5 \, 10^{-15} {\rm m}^{-2/3}$. Above 7km, the turbulence becomes very weak with $C_n^2$ values below $10^{-17} {\rm m}^{-2/3}$. At higher altitudes, above 18--20km, we observe sometimes an increase of $C_n^2$, due to the fact that the last points contain all the turbulent energy above, and that the reconstruction algorithm spreads this energy in the 2--3 upper levels.

Fig.~\ref{fig:cnmedian} (right) shows the fraction of the turbulent energy contained  between the ground and a given altitude $h$, defined as:
\begin{equation}
J(h)=\frac{\int_0^h C_n^2(h)\, dh}{\int_0^{h{\rm max}} C_n^2(h)\, dh}
\label{eq:J(h)}
\end{equation}
with $h_{\rm max}=24$km the highest altitude of PBL profiles. At night the first kilometer contains about 70\% of the turbulence. This is slightly greater than the value of 50\% observed at Paranal\cite{Osborn18}, but the Calern seeing is greater than the Paranal seeing. 

\begin{figure}
\includegraphics[width=15cm]{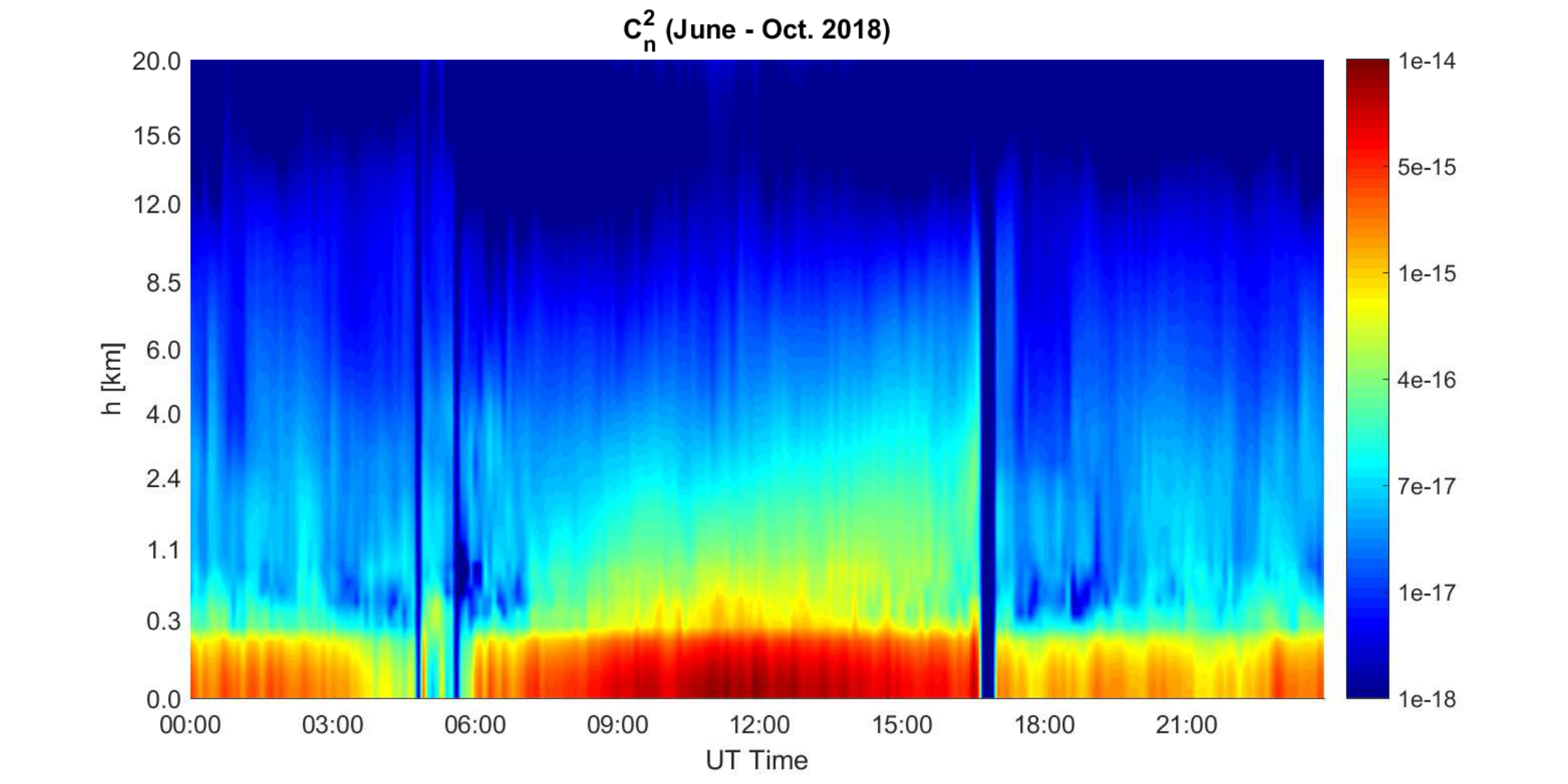}\\ \vskip -8mm
\includegraphics[width=14cm]{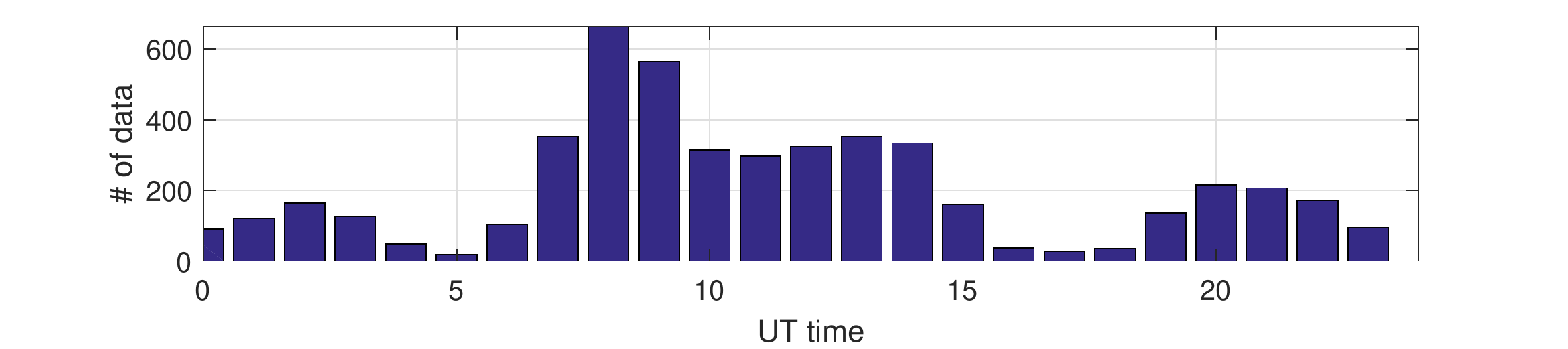}
\caption{Top: 24-hour evolution of the $C_n^2(h)$ profile for the period June-Oct 2018. Data are computed as the geometrical mean of individual profiles within time bins of 6 minutes. Bottom: number of recorded profiles per hour.}
\label{fig:cn2d24hr}
\end{figure}

During the day, the heating of the ground by the Sun creates an active turbulent layer is the first hundreds of meters. The first kilometer contains more of 80\% of the turbulent energy ($\sim$90\% at noon). This strong boundary layer is evident on the graph of Fig.~\ref{fig:cn2d24hr}. This curve was calculated as the geometric mean of individual profiles within time bins of 6~minutes. The ground turbulence is very strong between 7am and 4pm UT, with a maximum around noon. At nighttime, there is a sharp limit and the boundary layer seems to have a thickness of a little less than 300m. This boundary layer extends progressively during the day up to an altitude of 3--4km, until the sunset. The bottom panel of Fig~\ref{fig:cn2d24hr} shows the number of recorded profiles per hour: one can see that we lack observations during the transitions between the day and the night; more statistics will be needed to understand in details what happens.

\begin{table}
\begin{center}
\begin{tabular}{l|cc|cc} \hline
       &  $\epsilon$ (day) & $\epsilon$ (night) & $\theta_0$ (day) & $\theta_0$ (night)\\ \hline
Median & 2.07  &  0.97  &  1.69  & 2.15\\
First quartile & 1.55  &  0.76  &  1.31  & 1.72\\
Third quartile & 2.70 & 1.32 & 2.18 & 2.64 \\ \hline
\end{tabular}
\end{center}
\caption{PML statistics at Calern for the daytime and nighttime ground seeing $\epsilon$ and isoplanatic angle $\theta_0$ calculated at the wavelength $\lambda=500$nm. Values are in arcsec for the period June--October 2018.}
\label{table:statparams}
\end{table}

\begin{figure}
\begin{center}
\includegraphics[width=15cm]{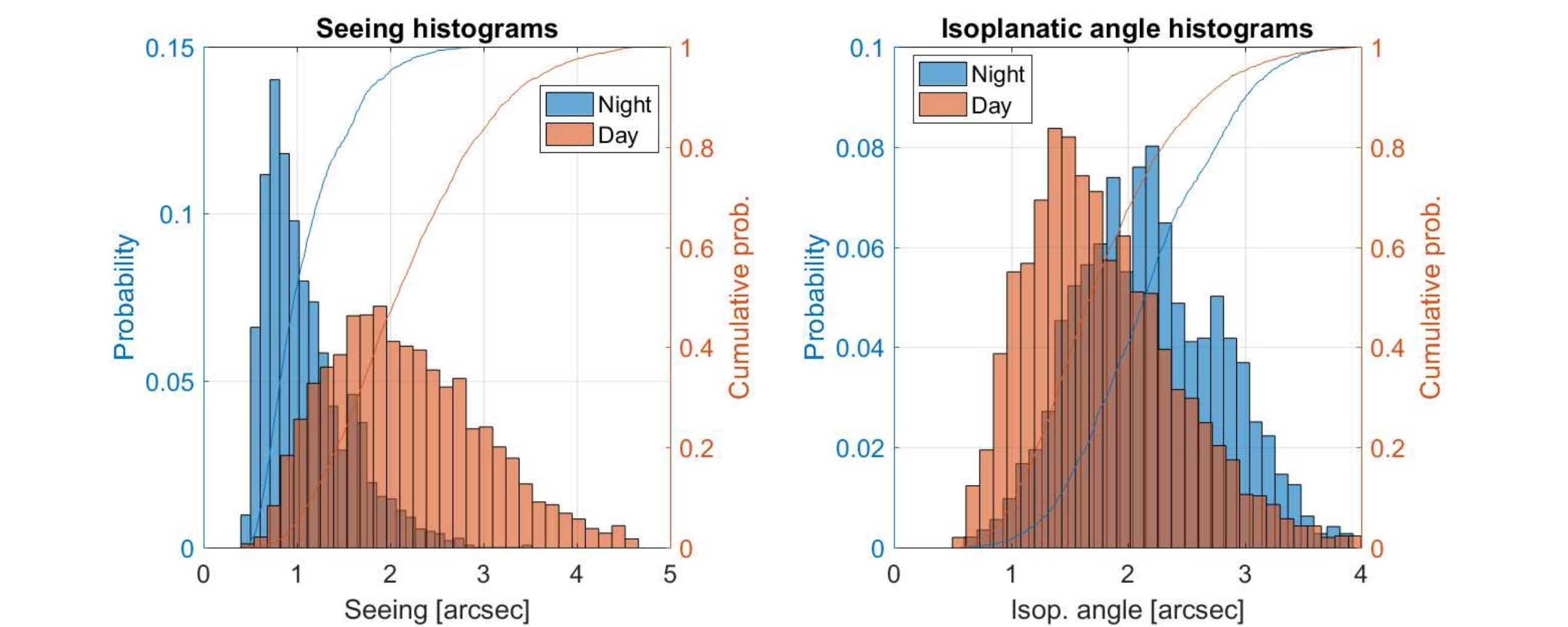}
\end{center}
\caption{Histograms of the daytime and nighttime seeing and isoplanatic angle at Calern during the period June--October 2018. Values are in arcsec for the wavelength $\lambda=500$nm.}
\label{fig:histparams}
\end{figure}

\subsection{Seeing and isoplanatic angle statistics}
The ground seeing is measured directly on lunar/solar limbs via the DIMM method. The isoplanatic angle makes use of the the $C_n^2$ profile via numerical integration of Eq.~\ref{eq:isop}. Statistics for both parameters at daytime and nighttime is summarized in Table~\ref{table:statparams}. Histograms and cumulative distributions are shown in Fig.~\ref{fig:histparams}. We observe the classical log-normal histograms, with an horizontal shift between the day and the night. With a median nighttime seeing of  $\sim 1''$, Calern appears to be an average site for high angular resolution astronomy. During the day, the measured seeing of $\sim 2$~arcsec is typical of values observed at the majority of solar sites (see for example the results of the site survey for the Advanced Technology Solar Telescope\cite{SocasNavarro05}).

\begin{figure}
\begin{center}
\includegraphics[width=14cm]{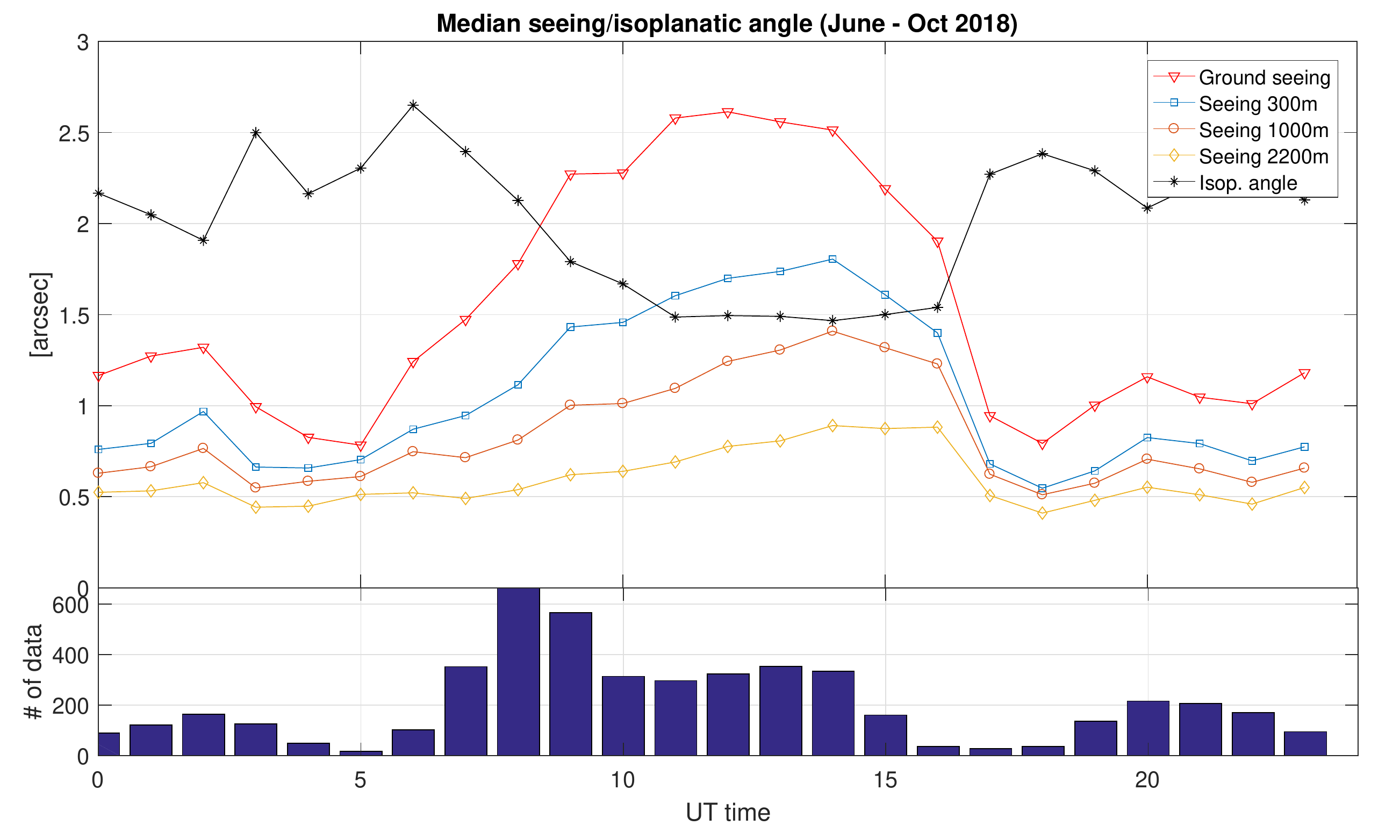}
\end{center}
\caption{Top: 24-hour evolution of the seeing at different altitudes and of the isoplanatic angle (values  recorded during the period June--Oct. 2018 for the wavelength $\lambda=500$nm). Bottom: number of recorded profiles per hour.}
\label{fig:seeingisopvstime}
\end{figure}

It is possible to obtain the seeing $\epsilon(h)$ as seen at an altitude $h$ above the ground, using  measured $C_n^2$ profiles, by numerical integration of the following integral\cite{Roddier81}
\begin{equation}
\epsilon(h)=5.25 \: \lambda^{-1/5}\: \left[\int_h^{h{\rm max}} dh'\, C_n^2(h')
\right]^{3/5}
\end{equation}
Fig.~\ref{fig:seeingisopvstime} shows the median hourly seeing measured at altitudes $h=0$m (ground), 300m, 1000m, and 2200m above ground level (agl) during the period June--Oct. 2018. The ground seeing is the most dependent of the time of the day, showing an increase of a factor 2 between the night and the middle of the day. This time dependence decreases with the altitude, but is still perceptible at $h=2200m$: the boundary layer extends over 2km. At night, the seeing above $h=300$m remains close to 0.6 arcsec and shows little dependence with time, this is typical to free atmosphere conditions.

The isoplanatic angle is also drawn in Fig.~\ref{fig:seeingisopvstime}, it shows nighttime values of $\sim 2''$ and a slight decrease to $\sim 1.5''$ during the day (from 9am to 4pm). This decrease is due to  high altitude turbulent layers which form during the day and disappear at night.

\begin{figure}
\begin{center}
\includegraphics[width=14cm]{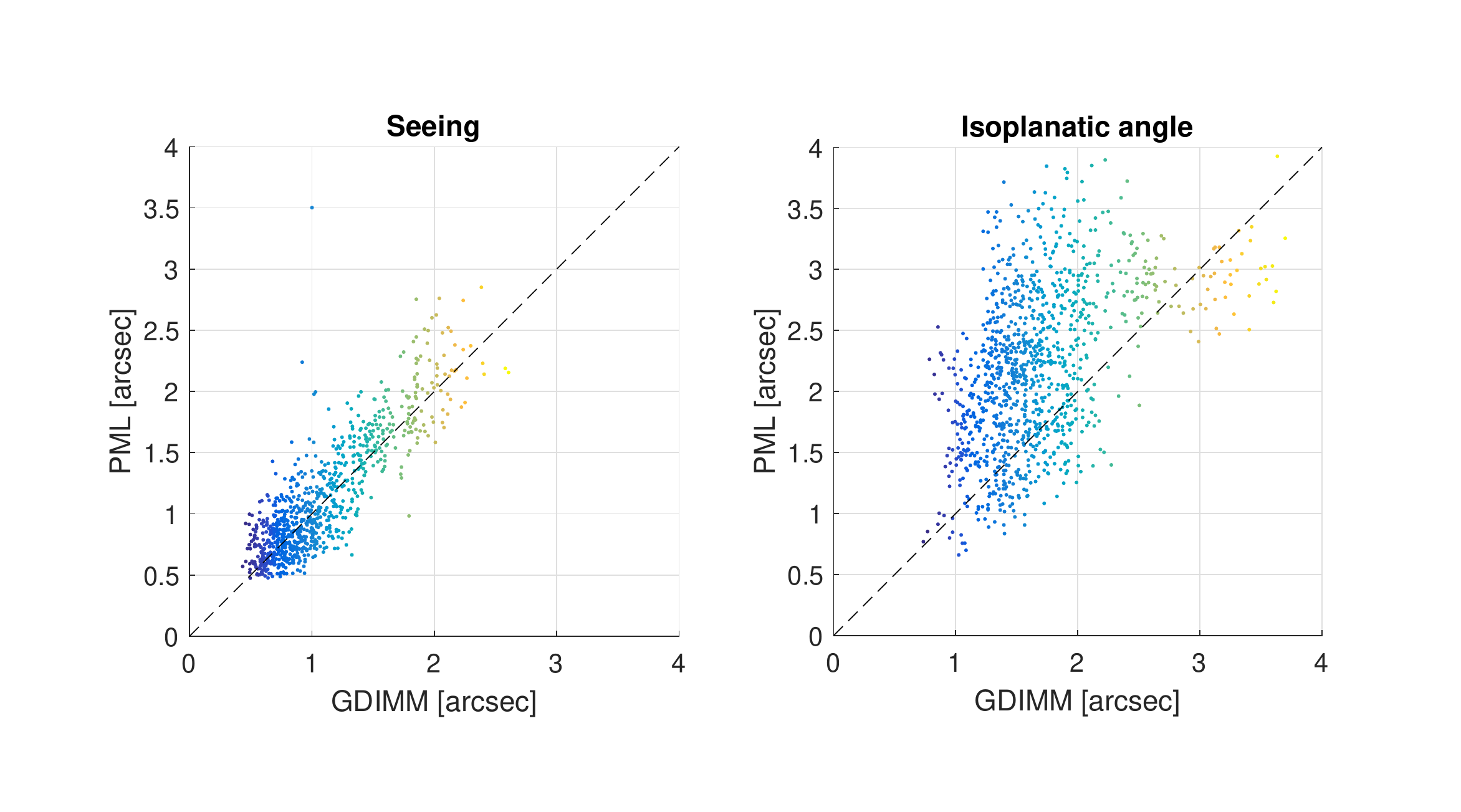}
\end{center}
\caption{Left: scatter plot of the nighttime seeing measured simultaneously by PML and GDIMM during the period June--Oct. 2018 (the wavelength is $\lambda=500nm$). Right: the same for the isoplanatic angle.}
\label{fig:coocgdimmpml}
\end{figure}

Simultaneous observations with the GDIMM monitor\cite{Aristidi19} were made during the night. GDIMM measures integrated parameters, among which the seeing and the isoplanatic angle. A comparison with PML is shown in the scatter plot of Fig.~\ref{fig:coocgdimmpml}. Each point of the plot corresponds to matching measures within a 1-minute time interval. For better comparison, GDIMM values were calculated for an exposure time of 5ms, corresponding to the PML exposure time. The leftern graph (seeing) shows a very good agreement between the two data sets, with a correlation coefficient of 85\%. The altitude difference of 4m between the two instruments is not a problem: with a ground $C_n^2\simeq 2. 10^{-15}$m$^{-2/3}$ at night, this 4-m thick layer contributes to less than 0.1 arcsec to the seeing and is below the error bars.
The scatter plot for the isoplanatic angle $\theta_0$ (Fig.~\ref{fig:coocgdimmpml}, right) shows a clear trend, but is somewhat more dispersed. This was remarked and discussed in [\cite{Ziad19}]. Reasons for these discrepancies are multiple. GDIMM's estimations of $\theta_0$ are made via the scintillation of a star, using a model\cite{LoosHogge79}. The scintillation is likely to be biased by thin clouds and/or moonlight (PML observations are only possible when the Moon is present). This results into an under-estimation of the isoplanatic angle by GDIMM. But a dedicated study of these bias effects in GDIMM measurements is needed before we can draw relevant conclusions.

\section{Conclusion}
The PML monitor is operational since 2016 as a part of the CATS station. It is now a fully automatic instrument, allowing continuous monitoring of turbulence profiles above the Calern observatory, daytime and nighttime. Data are displayed in real time through the website {\tt cats.oca.eu}. $C_n^2(h)$ profiles are calculated on a 33 levels altitude grid, with a vertical resolution fine enough to infer the properties of the boundary layer. The data processing pipeline has been optimized to allow real-time calculation of profiles which are available every 3~minutes. The instrument gives also access to the seeing and isoplanatic angle, results are consistent with other monitors such as the GDIMM.

Some results of the monitoring during the summer and autumn 2018 are presented. They show a strong difference between the day and the night. Nighttime conditions are typical of an average mid-latitude site with a seeing $\epsilon\sim 1''$, a ground $C_n^2\sim 2. 10^{-15}$m$^{-2/3}$, and a boundary layer of thickness $\lesssim 300$m. During the day, strong turbulence develops. The ground $C_n^2$ is multiplied by $\sim 10$ at noon and the seeing becomes greater than $2$arcsec. The boundary layer upper limit extends over 2000m. This is typical of daytime turbulence conditions measured at solar sites.

The PML has also the potentiality to estimate the profile of the wavefront coherence outer scale $\lo(h)$. The technique was developed some years ago for the Monitor of wavefront Outer Scale Profiles\cite{Maire07} and is based on the angular covariance of AA over a single Moon/Sun limb. The transposition of the technique to the PML is currently an ongoing process. 


\section*{acknowledgments}
%
%
We would like thank the Calern technical staff, for their valuable help on the instrument. The CATS project has been done under the financial support of CNES, Observatoire de la C\^ote d'Azur, Labex First TF, AS-GRAM, Federation
Doblin, Universit\'e de Nice-Sophia Antipolis and R\'egion Provence Alpes
C\^ote d'Azur. 

\bibliography{biblio}   
\bibliographystyle{spiebib}   

\end{document}